\documentclass{article}

\usepackage{arxiv}

\usepackage[utf8]{inputenc} 
\usepackage[T1]{fontenc}    
\usepackage{hyperref}       
\usepackage{url}            
\usepackage{booktabs}       
\usepackage{amsfonts}       
\usepackage{nicefrac}       
\usepackage{microtype}      
\usepackage{lipsum}		
\usepackage{graphicx}
\usepackage{natbib}
\usepackage{doi}

\usepackage{amsmath,amssymb,amsfonts}
\usepackage[binary-units=true]{siunitx}
\DeclareSIUnit \voltampere {VA} 
\DeclareSIUnit \var {VAr} 
\DeclareSIUnit \kWp {kWp} 
\DeclareSIUnit \kWh {kWh} 
\DeclareSIUnit \kWhperkWp {kWh/kWp} 
\DeclareSIUnit{\minute}{min}

\title{Hardware-Based Microgrid Coupled to Real-Time Simulated Power Grids for Evaluating New Control Strategies in Future Energy Systems}

\author{Michael Kyesswa\\
	Institute for Automation and Applied Informatics\\
	Karlsruhe Institute of Technology\\
	Karlsruhe, Germany \\
	\texttt{michael.kyesswa@gmail.com} \\
	\And
	\href{https://orcid.org/0000-0001-6984-7208}{\includegraphics[scale=0.06]{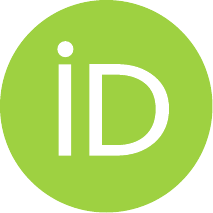}\hspace{1mm}Friedrich Wiegel} \\
	Institute for Automation and Applied Informatics\\
	Karlsruhe Institute of Technology\\
	Karlsruhe, Germany \\
	\texttt{friedrich.wiegel@kit.edu} \\
	\And
	\href{https://orcid.org/0000-0002-1194-6796}{\includegraphics[scale=0.06]{orcid.pdf}\hspace{1mm}Jan Wachter} \\
	Institute for Automation and Applied Informatics\\
	Karlsruhe Institute of Technology\\
	Karlsruhe, Germany \\
	\texttt{jan.wachter@kit.edu} \\
	\And
	\href{https://orcid.org/0000-0002-2218-3229}{\includegraphics[scale=0.06]{orcid.pdf}\hspace{1mm}Uwe Kühnapfel} \\
	Institute for Automation and Applied Informatics\\
	Karlsruhe Institute of Technology\\
	Karlsruhe, Germany \\
	\texttt{uwe.kuehnapfel@kit.edu} \\
	\And
	\href{https://orcid.org/0000-0002-9064-0196}{\includegraphics[scale=0.06]{orcid.pdf}\hspace{1mm}Simon Waczowicz} \\
	Institute for Automation and Applied Informatics\\
	Karlsruhe Institute of Technology\\
	Karlsruhe, Germany \\
	\texttt{simon.waczowicz@kit.edu} \\
	\And
	\href{https://orcid.org/0000-0002-3572-9083}{\includegraphics[scale=0.06]{orcid.pdf}\hspace{1mm}Veit Hagenmeyer} \\
	Institute for Automation and Applied Informatics\\
	Karlsruhe Institute of Technology\\
	Karlsruhe, Germany \\
	\texttt{veit.hagenmeyer@kit.edu} \\
}

\renewcommand{\shorttitle}{Hardware-Based Microgrid Coupled to Real-Time Simulated Power Grids}

\hypersetup{
pdftitle={Hardware-Based Microgrid Coupled to Real-Time Simulated Power Grids for Evaluating New Control Strategies in Future Energy Systems},
pdfauthor={Michael Kyesswa, Friedrich Wiegel, Jan Wachter, Uwe Kühnapfel, Simon Waczowicz, Veit Hagenmeyer},
pdfkeywords={Microgrids, Power grids, Power hardware-in-the-loop, Real-time simulations},
}

\begin{document}
\maketitle

\begin{abstract}
	The design of new control strategies for future energy systems can neither be directly tested in real power grids nor be evaluated based on only current grid situations. In this regard, extensive tests are required in laboratory settings using real power system equipment. However, since it is impossible to replicate the entire grid section of interest, even in large-scale experiments, hardware setups must be supplemented by detailed simulations to reproduce the system under study fully. This paper presents a unique test environment in which a hardware-based microgrid environment is physically coupled with a large-scale real-time simulation framework. The setup combines the advantages of developing new solutions using hardware-based experiments and evaluating the impact on large-scale power systems using real-time simulations. In this paper, the interface between the microgrid-under-test environment and the real-time simulations is evaluated in terms of accuracy and communication delays. Furthermore, a test case is presented showing the approach's ability to test microgrid control strategies for supporting the grid. It is observed that the communication delays via the physical interface depend on the simulation sampling time and do not significantly affect the accuracy in the interaction between the hardware and the simulated grid.
\end{abstract}


\section{Introduction} \label{sec:1Introduction}

The recent transformations in the power system structure have led to tremendous challenges in the management and control of power grids. These developments are mainly driven by the need to ensure a reliable and affordable future low-carbon power generation, which---according to recent studies~\cite{REN2018, IRENA2019}---will greatly depend on renewable energy sources (RES) such as solar photovoltaic and wind power generation. 
In terms of system properties, RES are typically interfaced to the grid through power electronics, unlike conventional generation sources which are interfaced through synchronous generators. These differences in the properties of the conventional generation sources and the RES result in challenges in system control from the operation point of view. For example, large rotating machines are the main sources of system inertia, which determines the rate of change of frequency (RoCoF). Therefore, replacing these machines with power electronic based sources will result in a reduction of such inherent grid services. In addition, the future grid faces another challenge due to the variability and intermittent nature of the renewable energy sources. One of the solutions to this challenge is to combine energy storage solutions and flexibility on the demand side of the grid in order to ensure a balance between generation and demand and thus stable grid operation. This has led to the concept of distributed energy resources (DER) in a smart grid context, considering the combination of renewable sources and the flexibility solutions. 

From the control point of view, the existing control architecture of the power grid with its various levels of control is tailored to the physically determined properties of the generation systems based on rotating machines. 
These characteristics are not inherent in power electronic converter based systems, and therefore control of such systems needs to be adjusted. 
The possible solutions that have been considered to address these challenges include integration of DERs with grid supporting services such as active power reduction for frequency regulation, or reactive power injection for voltage support to compensate for the reduced natural grid services~\cite{LMeegahapola2021}. 
While implementation of such control functions can potentially support the grid, integration of a large number of individual DERs creates additional challenges in terms of operation and control~\cite{RHLasseterMicrogrid2004}. The concept of microgrids 
is seen as a way of supporting the integration of DERs to the grid by considering a group of DERs and interconnected loads as a single controllable entity with respect to the grid~\cite{SParhiziMicrogrid2015, MFarrokMicrogrids2020}. 
They can be operated as AC microgrids, DC microgrids~\cite{JJacksonACDCmicrogrid2013}, or hybrid AC/DC microgrids~\cite{UEnekoHybridMicrogrid2015, AGupta2018}. 
From the grid perspective, microgrids enable decentralized coordination of DERs thus reducing the centralized control burden on the grid~\cite{NHatzMicrogrids2007}. The individual microgrids must therefore ensure compliant control of the DER components within, in order to provide the desired behavior in the overlying grid context. 


Research has widely been undertaken into development and implementation of microgrid technology ranging from microgrid architecture to control strategies. In~\cite{SParhiziMicrogrid2015}, a review of the state-of-the-art of microgrids research is given regarding the design, control and operation of microgrids both in grid-connected operation and in island operation mode. References~\cite{MLubnaReviewuGrid2013, MLubnaMicrogrid2016} give a review of existing microgrid architectures (including simulated microgrids) and highlight the benefits of grid-connected or island microgrids with energy storage systems. Further research has been undertaken in the development of control strategies for operation of microgrids as described in~\cite{ZRamonuGridControl2010, ABidram2012, DEOlivaresTrendsuGrid2014, TMorstynuGrid2018}. 
A simulation of coordinated control of a hybrid AC/DC microgrid is proposed in~\cite{TMaControluGrids2015}, considering grid-connected and island microgrid operation. Power management is another aspect that has been considered in microgrid control. Reference~\cite{MHosseinRobustuGrid2015} describes a power management system in hybrid AC/DC microgrids, whereby the power flow problem in the microgrid is formulated as a mixed integer linear programming problem. Other power management schemes have been presented for control of hybrid AC/DC microgrids in grid-connected operation during transient behavior as in~\cite{MAybaruGrid2021}.

However, the evaluation of new microgrid control solutions cannot be carried out directly in public grids in order to prevent interruption in grid operation and for safety of personnel and equipment. Therefore, rigorous experiments are required before implementation of the solutions in real power grids. Such experiments cannot be carried out only in terms of simulations but also require application of real laboratory hardware to test concepts in a hardware setting and integrate microgrids in large scale grid studies. Towards this end, the research on microgrids has also been supported in terms of real world grid infrastructure through demonstration projects and at the laboratory level as described in~\cite{NHatzMicrogrids2007, NWALidulaMicrogrids2011}. 
Reference~\cite{AKhurramDER2022} presents a real-time grid and DER cyber-physical co-simulation platform that can be used for testing DER coordination schemes. 
In addition, a recent concept of smart microgrids is increasingly considered in the current distribution systems in order to manage distributed resources and efficiently control bidirectional flow of electricity as experienced in the current power grids~\cite{MAybaruGrid2021}. 

The microgrid concept fits into the smart grid concept for testing demand side management and flexibility solutions. A number of smart grid infrastructures have been set up for research towards management of DERs and development of new solutions for future energy distribution systems as summarized in~\cite{ECommisionReport2018, LJansen2020}. Other smart  grid platforms have been set up to directly test and experiment innovative solutions in real power systems. Such laboratories include: the Energy Systems Integration Facility (ESIF) by the National Renewable Energy Laboratory; the Flex Power Grid Lab (FPGLab) in the Netherlands and the Concept Grid by \'Electricit\'e de France (EDF), among others. 

The available data about the hardware setting in the existing infrastructures shows a low degree of automation in the setup and supervision of microgrid experiments. This means a large share of time consuming manual intervention to reliably and safely operate the microgrid infrastructure. The facilities also lack concepts of legitimate generalization of component specific functions which is necessary to increase ease of setting up microgrid experiments. Furthermore, the physical connection within the facilities are limited to a low degree of freedom in terms of topology variation of the available microgrid components without time-consuming and costly manual intervention by experienced personnel. From the grid simulation side, the research infrastructures show experimental setups with simplified power grids and standard test networks.

The present paper addresses the above limitations with two novel experimentation approaches. On one hand, a novel framework is presented for setting up topologically variable hardware-based microgrid experiments in an automated and user-oriented research framework. With this, a framework is set up for modeling, control and validation of technologies for future energy systems in an experimental microgrid under test (MUT) setting. This provides a unique platform for investigating the interplay between microgrids and macrogrids, both grid-connected and island operation modes, as well as the transition between the two operation modes. It also allows the development and evaluation of new operational concepts for microgrids in the context of smart grids. On the other hand, since it is not possible to test the interaction between microgrid experiments and the main grid in the public grid, a unique large-scale simulation framework is presented for digital replications of real power grid models. This framework can be used to test software and hardware solutions for future grids. Furthermore, the paper presents an interface between the two frameworks, allowing for the advantages of developing control methods using real-part experiments and evaluating them in a system-wide setup with digital power grid models.

The aim of the present paper is to introduce the two frameworks and evaluate the physical interface between real-time simulated power grids and microgrid experiments set up using actual power system hardware. Furthermore, microgrid control strategies are described and evaluated in terms of supporting the main grid during disturbances. The main contributions of the paper are: 
\begin{itemize}
\item A novel topologically adjustable hardware-based microgrid framework in an automated and user-oriented research setting; 
\item A unique framework for accurate digital representation of real power grids in form of real-time simulated grids for application in experimental tests; 
\item The physical interface between real-time simulated power grid models and hardware-based microgrid experiments; 
\item A test scenario to demonstrate the application of the presented approach by testing a droop control strategy in the hardware-based microgrid DER setup for supporting the main grid during disturbances. 
\end{itemize}

The remainder of this article is organized as follows: Section \ref{sec:2sescl} gives a detailed description of the hardware-based microgrid setup provided by the Smart Energy System Control Laboratory. The unique simulation framework under the Energy Grids Simulation and Analysis Laboratory is described in Section \ref{sec:3egsal}. In Section \ref{sec:4interface}, the interface between the real-time simulated network and the hardware-based experiments in SESCL is described, together with the implemented control strategy in the microgrid experiment. The communication interface and control strategy are evaluated in Section \ref{sec:5results} with specific use case scenarios. Section \ref{sec:6conclusion} highlights the main conclusion of the paper, as well as an outlook into the future work.

\section{Microgrid Setup in the Smart Energy System Control Laboratory}\label{sec:2sescl}

\begin{figure}[!t] 
\centering
\includegraphics[width=0.5\textwidth]{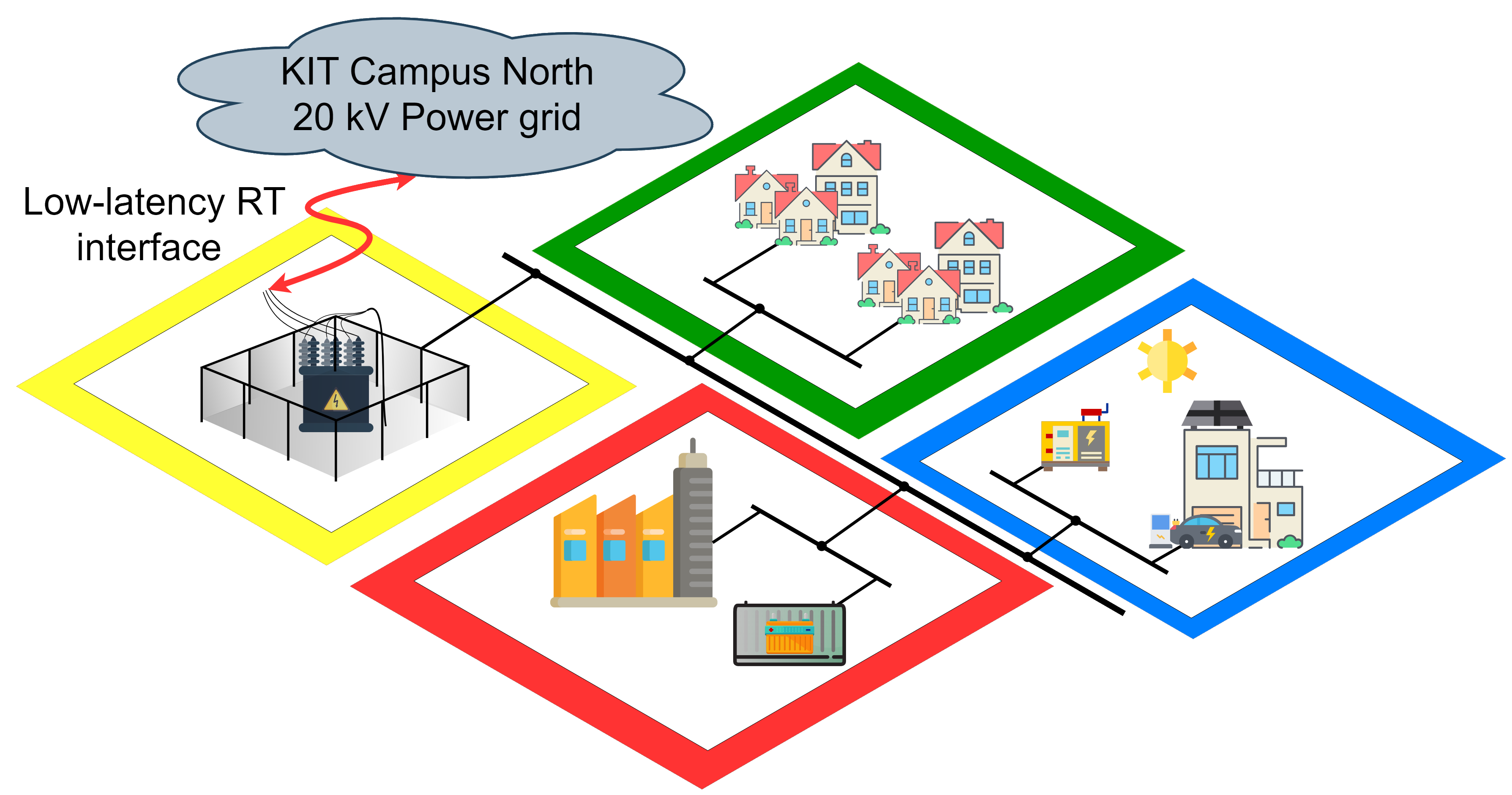}
\caption{Microgrid-under-Test (MuT) topology for interface to real-time simulated power grids. Green block is behavior replication of controlled RLC loads; Blue block are the physical components in SESCL; Red block are emulated components through PHIL; Yellow block is the interface to the simulated power grid}
\label{fig:fig_MuT}
\end{figure}

The microgrid setup described in the present article is provided, operated, and monitored by the Smart Energy System Control Laboratory (SESCL). SESCL’s high level of automation and ability to operate completely in a grid-decoupled way allow the study and assessment of tools and algorithms for energy technologies and grid control strategies on the edge of system stability in a secure setting. This attribute is essential since the energy transition cannot be designed in software simulations alone due to modeling shortcomings, and not all new ideas can be tested immediately in the public grid, due to a very high risk of a large-scale blackout. To provide a real grid environment, the laboratory integrates the most important equipment and machines as well as devices of the low voltage (LV) distribution grid, which can be interconnected fully automatically on demand and under desired topological specifications~\cite{wiegel_smart_2022}. However, this does not only consider distributed energy components, but also transmission line properties---via real transmission lines, cables, and replicas with adjustable lengths ---forming a physical link between the components. 
The present section describes the properties of the power system components in SESCL that form the case study microgrid as illustrated in Fig.~\ref{fig:fig_MuT}.

\begin{figure}[!b] 
\centering
\includegraphics[width=0.5\textwidth]{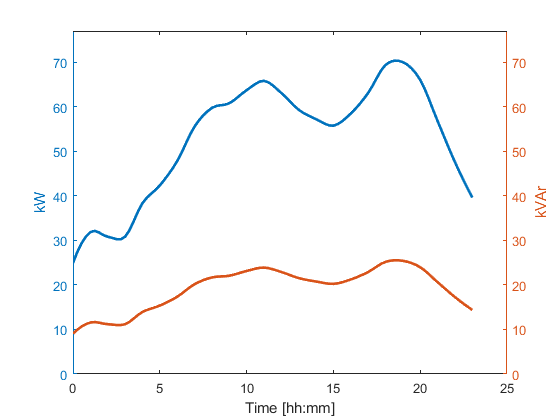}
\caption{Load profile of the residential block used in the case study}
\label{fig:fig_Dayprofile}
\end{figure}

The load behavior of a residential block represented in green in Fig.~\ref{fig:fig_MuT} is reproduced by controlled RLC loads, whereby only the RL elements are used for the current study case. These loads enable the provision of an adjustable passive PQ-Load in the range of $0 - 89$ \si{\kilo\watt} and $0 - 89$ \si{\kilo\var} for the active and the reactive power, respectively. The available step size corresponds to \SI{330}{\watt} resp. \si{\var}. 
The effective load profile applied for the experiment is represented in Fig.~\ref{fig:fig_Dayprofile}. The setup maximum load of \SI{78}{\kilo\voltampere} corresponds on average to approx. $20$ household units. Thereby the profile is derived from the aggregated measurement data of the three LLEC houses~\cite{Bottaccioli2019}. To minimize the phase imbalance and to simplify the consideration, the instantaneous reactive and active power are divided equally between the three phases. The average power factor (cos phi) is set to $0.9$. According to several recent studies, this pattern roughly corresponds to the load profile of a four-person household~\cite{Bottaccioli2019}.

The blue part of Fig.~\ref{fig:fig_MuT} is represented by real components within the SESCL. For this purpose, the heat pump (HP) house of the LLEC and the emergency generator are integrated into the study case microgrid. In addition to a heat pump, the HP house is equipped with a PV system, a Home Energy Storage System (HESS) and a wallbox for electric vehicles (EVs). The PV system with \SI{9.92}{\kWp} generator power and measured spec. annual yield \SI{818.15}{\kWhperkWp} is connected to HESS via the DC link. The HESS is the S10 Blackline system with \SI{10.56}{\kWh} from E3DC.

\begin{figure}[!t] 
\centering
\includegraphics[width=0.5\textwidth]{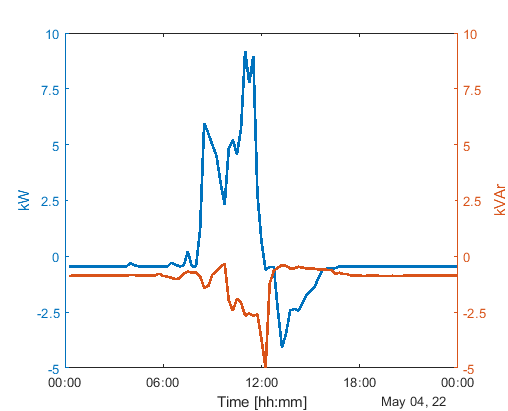}
\caption{Load profile of the heat pump (HP) house on the day of the case study}
\label{fig:fig_hp_dayprofile}
\end{figure}

Fig.~\ref{fig:fig_hp_dayprofile} shows a daily load profile---averaged over \SI{15}{\minute}---at the home's feed point on one of the experimental days.  The peak load of over \SI{9}{\kilo\watt} between 10.00 am and 12.30 am is due to a charging process of an electric vehicle at the wallbox of the house with a set charging power of \SI{7}{\kilo\watt}. Since it was relatively warm and very cloudy on this day, see Fig.~\ref{fig:fig_hp_dayprofile}, there is no influence of the heat pump and only a small power injection of the PV system between 01.00 pm and 05.00 pm can be seen on the daily profile.  

The emergency generator is a gas-powered combustion engine GM52 from Feeser Generators with a continuous output of \SI{56}{\kilo\voltampere}. In this test case, the emergency generator is configured that it only starts up and forms an island grid for the heat pump house in the event of a blackout of the general supply. Otherwise, the generator does not actively participate in the grid supply. 

The interface between SESCL's microgrid-under-test (MUT) and the real-time simulated power grids is performed by two-stage Real Time Simulator + Power-Hardware-in-the-Loop test environment (RTS-PHIL-TE) provided jointly by EGSAL and SESCL. Thereby, the behavior of distribution grid substation---yellow part of Fig.~\ref{fig:fig_MuT}---is mimicked with the help of the Ideal Transformer Method (ITM)~\cite{9862070}. Besides the distribution grid substation as an interface, the Battery Energy Storage System (BESS) in the red part of Fig.~\ref{fig:fig_MuT} is emulated by PHIL in the presented experiment to test the provision of grid-friendly services, such as the provision of active and reactive power. The detailed description of the used control algorithm as well as the realization of the interface is presented in Sec.~\ref{seq:MuT_KIT_CN}. In both cases, a PHIL setup consisting of OPAL-RT's OP5707 real-time simulator and a CSU100 2GAMP4 four-quadrant power amplifier from Egston is used for emulation. 

The electrical connection, i.e. the microgrid topology, is established via the busbar matrix composed of ten busbars and a total of $424$ contactors of different types and power classes~\cite{wiegel_smart_2022}. This setup allows the realization of load shedding or integration of additional generators, consumers or prosumers during runtime and changing the grid topology on demand and in real-time. The comprehensive automation system monitors and controls both the matrix and the individual members of the microgrid and the schedule of the experiment. It is responsible for recording, visualizing and logging all experimental, plant-related and component-specific measured quantities, such as current, voltage, mains frequency, power, harmonics, etc. The variables to be measured are captured in the AC domain of the matrix at \SI{20}{\kilo\hertz}~\cite{wiegel_smart_2022}.

\section{The Digital Framework in the Energy Grids Simulation and Analysis Laboratory}	\label{sec:3egsal}
 
The Energy Grids Simulation and Analysis Laboratory provides a digital framework for modeling, simulation, analysis and visualization of energy grids of different sizes and at different voltage levels. The main components of the laboratory that are relevant in the setup of the experiment introduced in the present paper are the digital power grid models, the real-time simulation and interface to external devices as described in the following sections.

\subsection{Power Grid Models}
Within the EGSAL simulation framework, the scope of power grid models ranges from low-voltage microgrids to very high voltage transmission networks. Grids of different sizes in the terms of structural complexity are considered during the modeling and simulation. Examples of the power grids available in the EGSAL framework include the KIT Campus North \SI{20}{\kilo\volt} power grid, distribution and transmission networks, such as the Karlsruhe city network, Baden-W\"{u}rttemberg $380/220/110$\,\si{\kilo\volt} network, the Germany $380/220$\,\si{\kilo\volt} network and the interconnected central European $380/220$\,\si{\kilo\volt} transmission grid. 

\begin{figure}[!h] 
\centering
\includegraphics[width=0.95\textwidth]{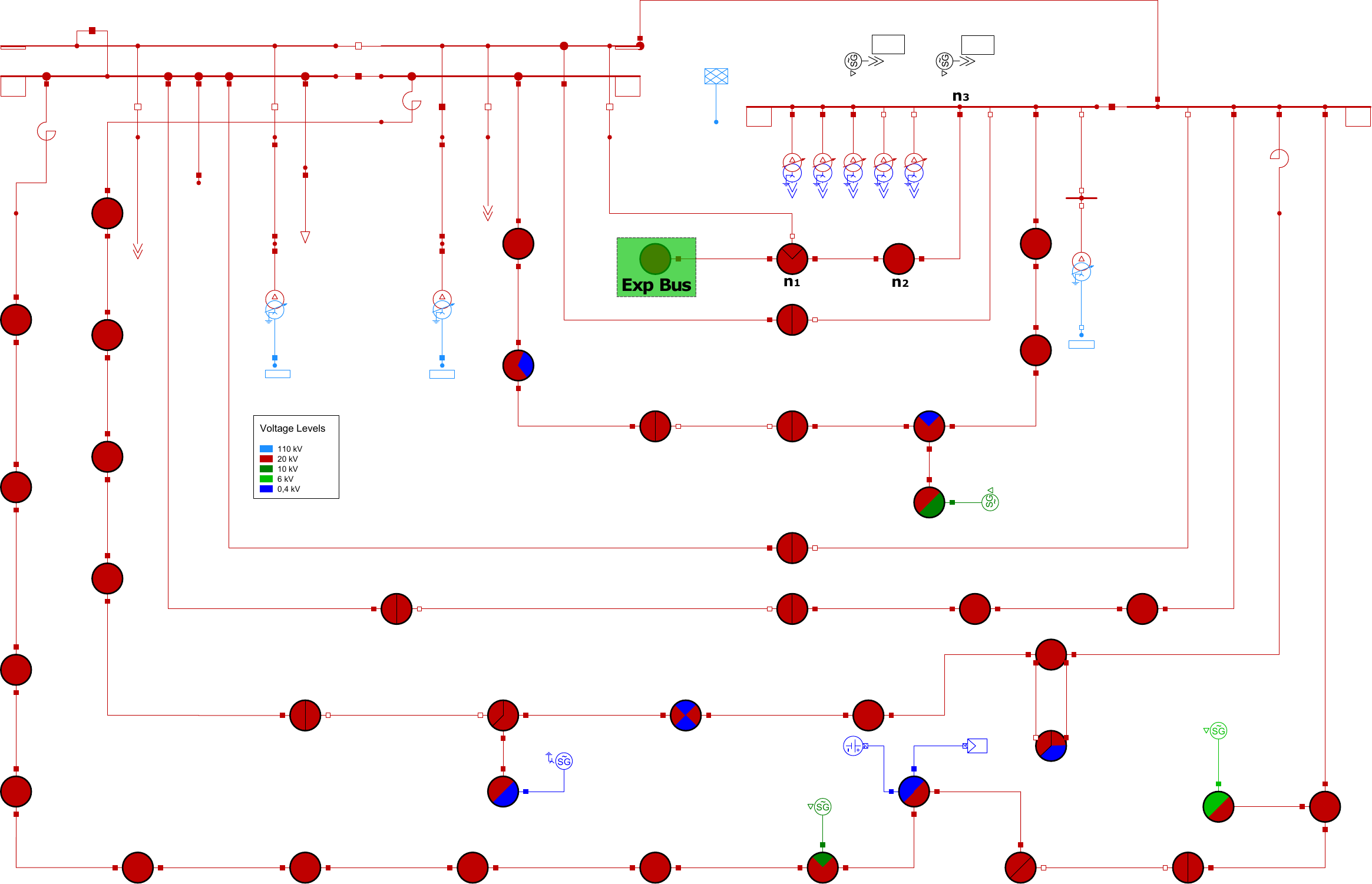}
\caption{KIT north campus \SI{20}{\kilo\volt} electrical network in PowerFactory. The coupling point of the experiment hall to the campus network is the \textit{Exp Bus} green highlighted block}
\label{fig:fig_KITCN}
\end{figure}

In the present paper, the grid applied in the experimental setup is the benchmark model of the KIT Campus North \SI{20}{\kilo\volt} network as described in \cite{MWeber2021}. Fig.~\ref{fig:fig_KITCN} shows the digital representation of the KIT Campus North network in DIgSILENT PowerFactory simulation software. The components in the network structure can be summarized as follows: 43 \SI{20}{\kilo\volt}-buses, 87 transformers, 86 loads connected to the \SI{400}{\volt} level,  four internal generators and one \SI{1.5}{\mega\watt} peak solar PV generation. Further details about the campus network and structural representation are given in \cite{MWeber2021}. 

\subsection{Real-time Simulation and External Interface}
EGSAL provides a variety of hardware and software tools for running simulations ranging from offline and parallel grid simulations to real-time simulations. The present paper is, however, limited to real-time simulations. The main components for the real-time simulations of electrical grids within the scope of EGSAL are the Real-Time Digital Simulators (RTDS). 

Interface of the simulators to external devices can be realized either through a direct interface or through a virtual connection. 
The different interface methods to other simulators and to external devices are described in the following.

\subsection*{Simulator -- Simulator Coupling}\label{subsec: rack connection methods}
The size of power grid in a real-time simulation is limited by the processing power of the simulator. In EGSAL, the RTDS Navacor chassis has 10 CPU cores with a limit of 300 load units. Interconnecting several simulators creates a larger simulation clusters, whereby the simulators combine their processing power and thus allows simulation of larger networks. The following simulator coupling methods have been tested within the framework of EGSAL:  

\subsubsection*{Inter-Rack Communication (IRC)}
This is a direct link between different simulators realized through the IRC connections on each RTDS rack (i.e. six IRC connections per RTDS) via a fiber optic cable. Additionally, an IRC switch can be used to augment the IRC connections in the inter-rack communication, thus extending the racks in the simulation cluster. 
Distribution of the network among the different simulators requires splitting the network into subsystems, which are interconnected using the transmission line model method. This method depends on the properties on the traveling electromagnetic wave and the simulation step size, i.e. if the traveling time is greater than the simulation step size, the changes at the one end of the transmission line or cable do not affect the other end of the line in the same time step. 

\subsubsection*{GTNETx2 Interface} 
\begin{figure}[!ht]
\centering
\includegraphics[width=0.65\textwidth]{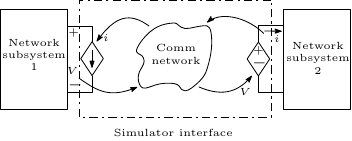}
\caption{Ideal transformer model interface method}
\label{fig:itm_method}
\end{figure}

The network interface card GTNETx2 is an interface between the real-time simulator and external devices. Using the GTNETx2 interface method, additional simulators in the simulation cluster are considered as external devices and the GTNETx2 card provides a communication link between the main simulator and the secondary simulators via Ethernet using standard communication protocols. Once the network is distributed among the available simulators, specific coupling points are defined for interfacing the different subsystems. The most common interface method in such a distributed real-time simulation is the Ideal Transformer Model (ITM) method \cite{WRen2008}. The coupling point of the distributed network is implemented as a controlled current source in one subsystem and a controlled voltage source in the other. The ITM interface method is illustrated in Fig.~\ref{fig:itm_method}. The communication bandwidth depends on the applied network protocol, e.g. the GTNET-SKT protocol supports 300 input and output data points (each with 4 bytes) per packet and 5000 packets per second.

\subsubsection*{Virtual Connection via VILLASnode:}
\begin{figure}[!h]
\centering
\includegraphics[width=0.65\textwidth]{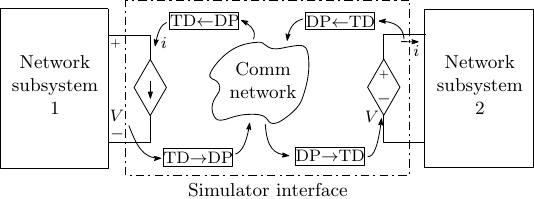}
\caption{Ideal transformer model with dynamic phasors interface method}
\label{fig:itm_dp_method}
\end{figure}

The virtual interface is used to extend the simulation cluster by creating a virtual interconnection to geographically distant external simulators. The special hardware used to realize this interconnection is a server running VILLASnode for geographically distributed real-time simulations \cite{SVogel2017}.
VILLASnode is a modular gateway application that is designed for low latency processing and forwarding of data to achieve real-time simulation \cite{VillasOnline2020}. The interface method in the virtual connection is based on the ideal transformer model method. 
The accuracy of the ITM method is, however, affected by delays and phase shifts between the measured voltage and current signals in the interfaced subsystems. For this reason, \cite{MStevic2015} proposes a modification of the ITM method by applying dynamic phasors as the exchanged values between simulators in the virtual interface. In comparison with Fig.~\ref{fig:itm_method}, Fig.~\ref{fig:itm_dp_method} illustrates the modification in the ITM method with dynamic phasors as the exchanged signals. The received dynamic phasors (DP) in the respective subsystems are converted back to time-domain (TD) representation.   

\subsection*{Interface to External Hardware}\label{subsec: external_interface}
The direct interface to external hardware can be realized either through the Gigabit Transceiver (GT) digital and analogue cards or through the link layer communication via Aurora protocol. 

\subsubsection*{Gigabit Transceiver digital/analogue cards}
Analogue signals from external devices are interfaced to the simulator via the Gigabit Transceiver Analogue Input Card. The analogue channels are sampled synchronously with new samples sent to the simulator every \SI{1}{\micro\second}. The Gigabit Transceiver Digital Input Card is used to interface digital signals from external devices to the RTDS simulator. 
Similarly, analogue and digital signals from the RTDS simulator are interfaced to external devices via Gigabit Transceiver Analogue Output card or Digital Output card, respectively. 

\subsubsection*{Link layer communication via Aurora protocol}
External devices can also be interfaced to the RTDS simulator using the Aurora 8B/10B protocol. This is a high-speed digital interface to the simulator via GT fiber cable. Four GT input/output ports on each RTDS simulator are reserved specifically for interfacing external hardware via Aurora communication. This implies that four simultaneous communication channels can be enabled on the simulator. The interface between the simulator and the external devices is realized using the ITM interface method (cf. Fig.~\ref{fig:itm_method}). 

\section{Microgrid and Simulated Campus Network Interface} \label{sec:4interface}
\label{seq:MuT_KIT_CN}

The present section describes the interface between the simulated power grid models and a hardware-based microgrid experimental setup using a Power-Hardware-in-the-Loop approach. The aim of these coupled parts is to develop and evaluate control strategies in energy systems using actual power system components and real-time representations of power grids. The KIT Campus North \SI{20}{\kilo\volt} power grid is used for the setup described in this paper. 

\subsection{Experimental Setup}
\label{sec:interface}
\begin{figure}[!h] 
\centering
\includegraphics[width=0.65\textwidth]{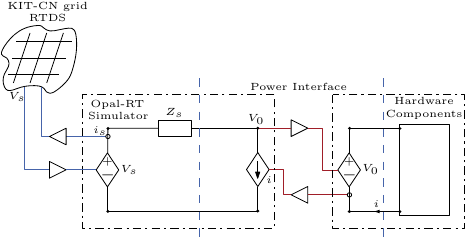}
\caption{Schematic representation of the experimental setup of the interconnection between RTDS KIT Campus North power grid and the OPAL-RT-PHIL experiment}
\label{fig:fig_opalRT_RTDS}
\end{figure}

The experimental setup consists of two digital real-time simulators, i.e. RTDS and OPAL-RT with a direct high performance interface using the Aurora protocol, connected to experimental hardware. This constitutes the PHIL setup as illustrated in Fig.~\ref{fig:fig_opalRT_RTDS}. 
The point of coupling of the two systems is at the \SI{400}{\volt}-bus to which the SESCL laboratory is physically connected in the KIT Campus North \SI{20}{\kilo\volt} network.
%
As shown in Fig.~\ref{fig:fig_opalRT_RTDS}, current and voltage signals are applied as the interface signals in the generic PHIL setup. On one side, a voltage signal is calculated from the OPAL-RT simulation and fed into the terminals of the real hardware experiment via a power interface. On the other side, current is measured from the experimental circuit and fed-back into the simulated model on the OPAL-RT system.

Similarly, current and voltage signals are used for the interface between the PHIL setup on OPAL-RT and the simulated campus network in RTDS. From the grid side in RTDS, the OPAL-RT-PHIL setup is seen as a current source, whereas the campus network is a voltage source in the OPAL-RT-PHIL setup. The exchange of the signals is via the Aurora 8B/10B communication protocol using Aurora links in each simulation model. In this case, the instantaneous node voltage at the coupling bus in the campus network in RTDS is sent to the OPAL-RT model as input to the voltage source, whereas current is measured in the OPAL-RT simulation model and sent to the RTDS simulation as input to the current source.
\subsection{Data Processing of the Received Three Phase Voltage}
As depicted in Fig.~\ref{fig:fig_opalRT_RTDS}, the OPAL-RT real-time simulator receives the three phase voltage at the coupling point via the previously described communication link. Due to the unavoidable loss of data during the transport process, the respective gaps need to be filled. As for the present implementation, no detection of data loss is possible and missing data is replaced by zeros in the voltage and thus corrupt the voltage signal. The direct use of the received voltages as setpoints for the corresponding amplifiers leads to severe issues. Due to the large bandwidth of the amplifier, the amplifier voltage follows the corrupted signal. 
To mitigate the issue in the presented experiments, the angle and amplitude information is extracted from the voltage signal. The angle information is obtained using a standard dq-transformation based Phase-Locked Loop (PLL), therefore no additional delay is introduced for the angle information. For the amplitude a moving average filter signal of the RMS value is used by choosing an appropriate window length, whereby the disruptive effects can be reduced while only introducing a small delay for the RMS value.
\subsection{Droop Control in Microgrid}
\begin{figure}[!h] 
\centering
\includegraphics[width=0.60\textwidth]{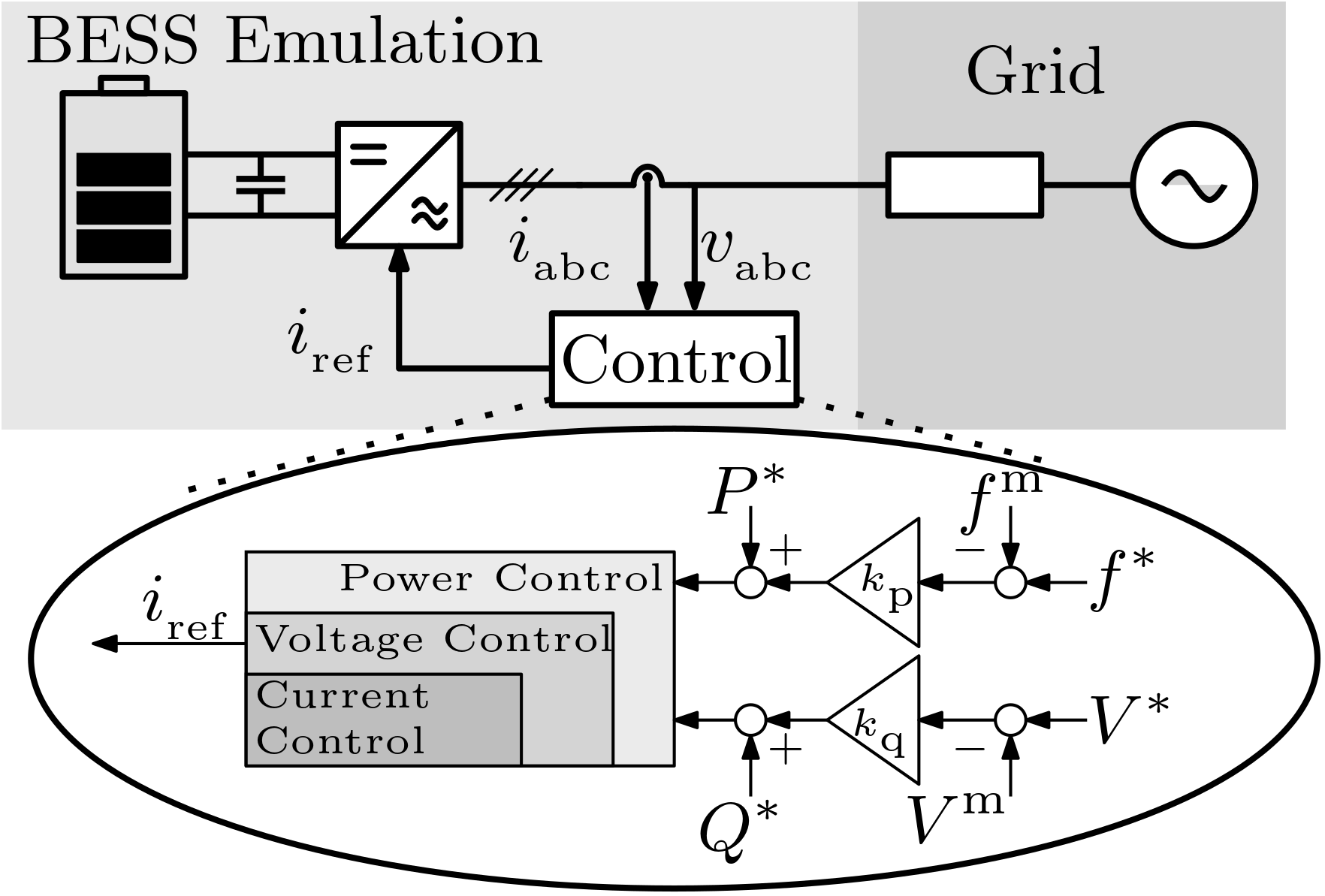}
\caption{Schematic representation of BESS control system}
\label{fig:fig_BESScontrol}
\end{figure}

For the control approach of the BESS in the microgrid, a droop strategy is used, which is a well known approach that is inspired by the natural behavior of synchronous generation~\cite{Sun2017,Tayab2017}. The advantages of droop control strategies lies in their simplicity and the fully decentralized scheme which does not rely on information from outside~\cite{Brabandere2007}. Furthermore, the design of the droop parameters allows for power sharing according to a desired ratio, e.g. the rated power. Since the aim in the present article is to introduce the interaction between the simulation and PHIL environment rather than the control approach, a simple active power--frequency (P/f), reactive power--voltage (Q/V) droop is used even though this is under the assumption of an inductive grid~\cite{Rocabert2012}. This implies a relation of the deviation in frequency to active power and the deviation in the voltage level with reactive power as follows
\begin{align}
    \label{eq:DroopEquations}
     P_\mathrm{D} &= k_\mathrm{P}\, (f^{*}-f_\mathrm{m}) \notag\\
     Q_\mathrm{D} &= k_\mathrm{Q}\, (V^{*}-V_\mathrm{m}),\notag
\end{align}
where $f^{*},\,V^{*}$ denote the nominal values for frequency and RMS Voltage of the grid, the corresponding measured values are given by $f_\mathrm{m},\,V_\mathrm{m}$, and $P_\mathrm{D},Q_\mathrm{D}$ are the resulting setpoint contributions for the underlying power control loop. The active and reactive power setpoints are $P^{*}$ and $Q^{*}$, respectively. Fig.~\ref{fig:fig_BESScontrol} depicts the control approach used in the present scenario. Note that due to the amplifier used for the BESS emulation, whereby the control of the amplifier is set to current control mode, the output of the implemented control is the desired current $i_\mathrm{ref}$.

\section{Results and Evaluation} \label{sec:5results}

The current section evaluates the interface between the simulated campus network and the microgrid experimental setup with real hardware components in SESCL. The point of coupling in this case is directly on the LV side of the transformer connecting the Energy Lab 2.0 to the campus grid in the simulated network. 
The purpose of the experiment is to evaluate the communication delays in the closed loop RTDS -- OPAL-RT -- PHIL setup and determine how such delays affect the simulation accuracy. Furthermore, the droop control function of the BESS in the microgrid setup is tested to analyze the influence of the BESS system on the main grid in terms of support during network disturbances. 



\subsection{Minimum Delay Test Case}\label{subsec:5_benchmark}
\begin{figure}[!t] 
\centering
\includegraphics[width=0.75\textwidth]{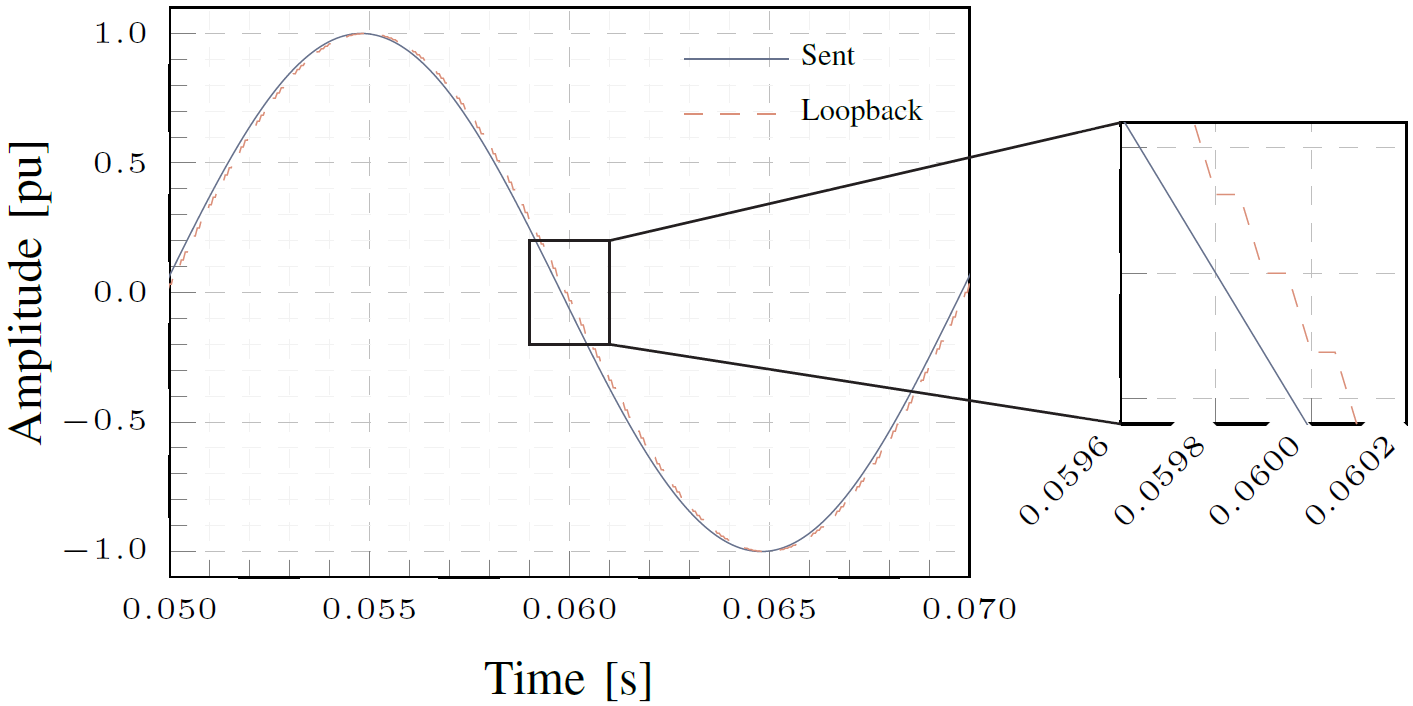}
\caption{Minimum delay measurement between the simulators via the high performance Aurora interface; Comparison of sent signal from RTDS to loopback signal from OPAL-RT}\label{fig:benchmark}
\end{figure}

Initially, the communication between the two simulators RTDS and OPAL-RT via Aurora protocol is tested to determine the minimum communication delay as a reference for the delay measurements in the experimental setup. The simulation time step on both simulators is \SI{50}{\micro\second}. A signal is sent from RTDS to OPAL-RT (\textit{sent}), and looped back to the RTDS simulator (\textit{loopback}). The two signals are recorded on the RTDS simulation side as shown in Fig.~\ref{fig:benchmark}. 

Analysis of the \textit{sent} and \textit{loopback} signals shows that the delay in the closed loop signal transmission is \SI{100}{\micro\second} as shown on the zoomed-in scale in Fig.~\ref{fig:benchmark}. This corresponds to two simulation time steps, i.e. one time step to send the signal from RTDS to OPAL-RT and one time step to loopback the signal from OPAL-RT to RTDS. From these results, it can be concluded that the link layer delay via Aurora protocol has negligible influence on the communication between the simulators. The resulting communication delay mainly depends on the simulation sampling time. 

\subsection{Campus Network -- Microgrid Experiment}\label{subsec:5_experiment}

In the following case, the interface between the microgrid and the campus network is tested by a set of events on the simulated network side and on the physical hardware side. The purpose of the experiments is to analyze the performance of the high performance physical interface following changes in the set points and whether each side of the experiment correctly responds to an event in the opposite side. The following scenarios are tested as described in the sections below: i) Sudden increase in load on the simulated grid side; ii) Unbalanced load change on the physical hardware side in the microgrid. 

\subsection*{Sudden Load Increase by \SI{1}{\mega\watt}}\label{subsec:5_exp1}

The first scenario evaluates the response of the hardware microgrid to an event in the simulated network. In this case, a sudden increase in load by \SI{1}{\mega\watt} is simulated at the point of common coupling on the network side. The key system parameters analyzed in this case include the grid frequency, the injected active power into the microgrid, voltage at the point of common coupling, and voltages at other measurement points close to the point of common coupling in the simulated network. The responses of the selected key parameters are compared for cases with (\textit{with droop}) and without (\textit{no droop}) droop control function in the BESS system. 

\begin{figure}[!t] 
\centering
\includegraphics[width=0.75\textwidth]{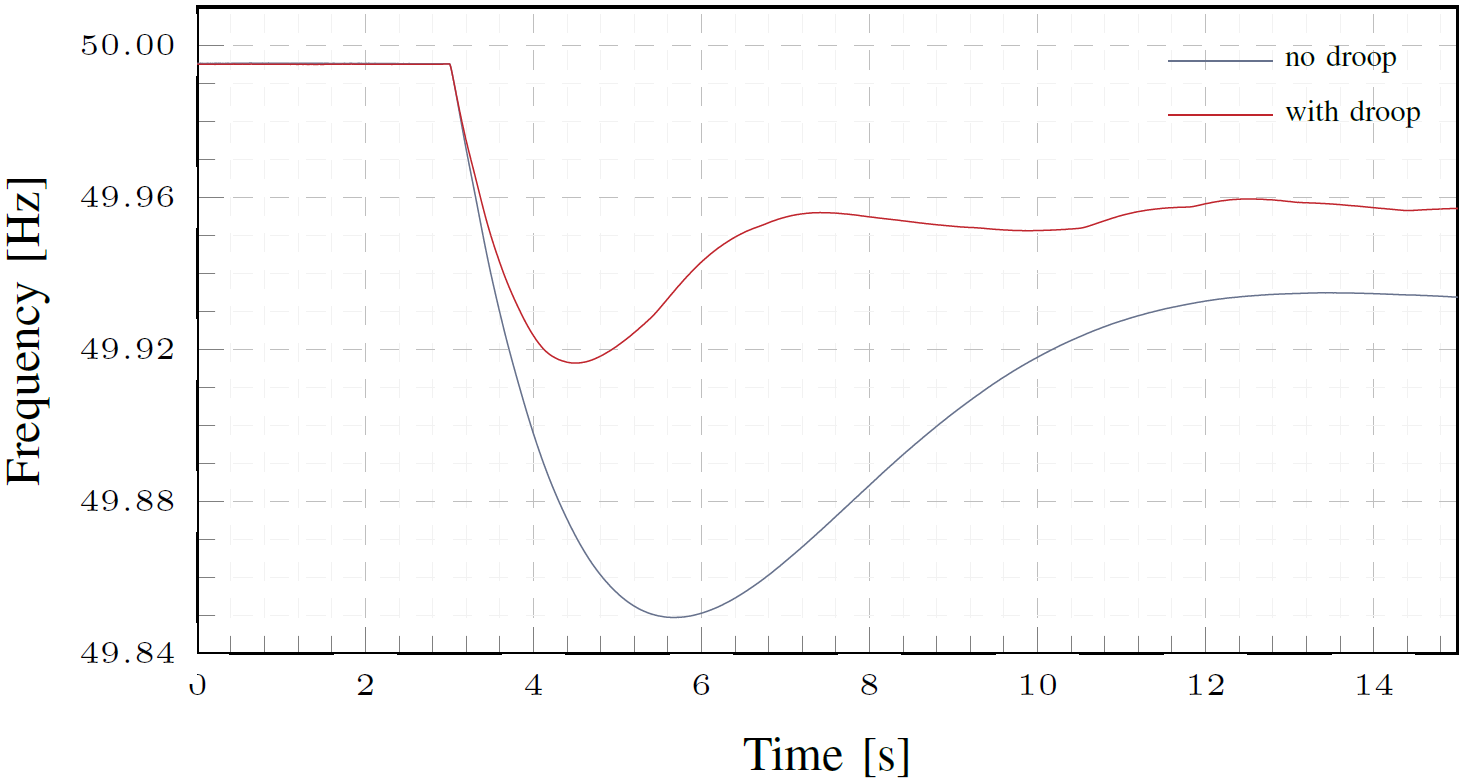}
\caption{Response of grid frequency following a load increase by \SI{1}{\mega\watt} with and without droop control of the BESS system in the microgrid}\label{fig:frequency}
\end{figure}

\begin{figure}[!t] 
\centering
\includegraphics[width=0.75\textwidth]{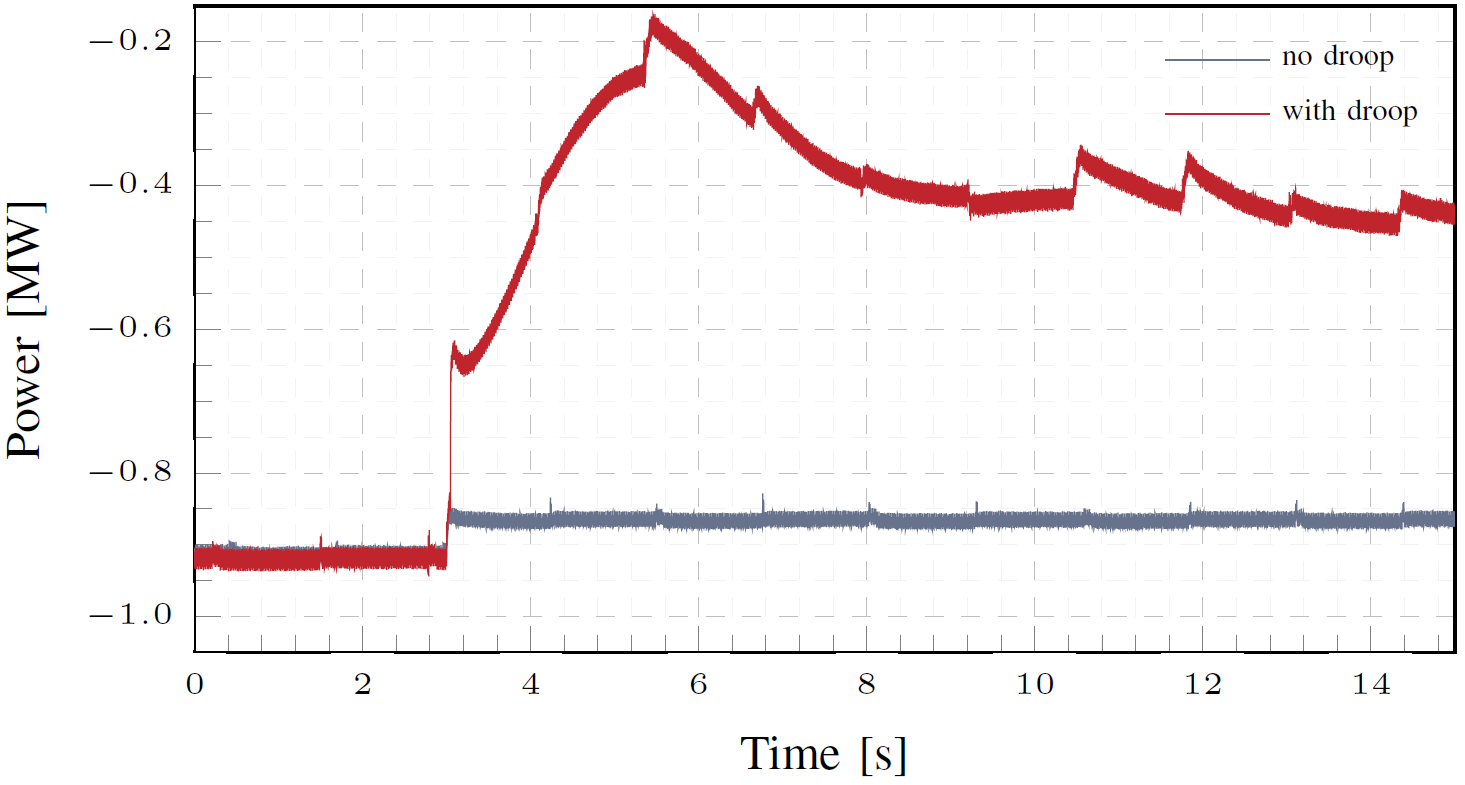}
\caption{Active power injection into the microgrid in response to a \SI{1}{\mega\watt} load increase. Negative power implies powerflow into the microgrid whereas positive power is powerflow from the microgrid}\label{fig:acitve_power}
\end{figure}

Fig.~\ref{fig:frequency} shows the grid frequency in response to the sudden increase in load for the two experimental scenarios with and without droop control from the BESS system. Important to observe is the additional frequency support from the droop control of the BESS system to stabilize the grid frequency closer to the acceptable range of operation. This is achieved by the BESS system injecting additional power in response to the load increase in the main grid. This in turn corresponds to a reduction in the active power absorbed by the microgrid from the main grid as shown in Fig.~\ref{fig:acitve_power}, compared to the power absorbed in the scenario without activation of droop control in the BESS system. The artifacts observed in the active power profile are a result of disturbances in the communication link between the real-time simulated network and the hardware microgrid experiment. Corresponding investigations will consider application of dynamic phasors to the exchanged signals as a potential solution to the communication problem. 


\begin{figure}[!t] 
\centering
\includegraphics[width=0.75\textwidth]{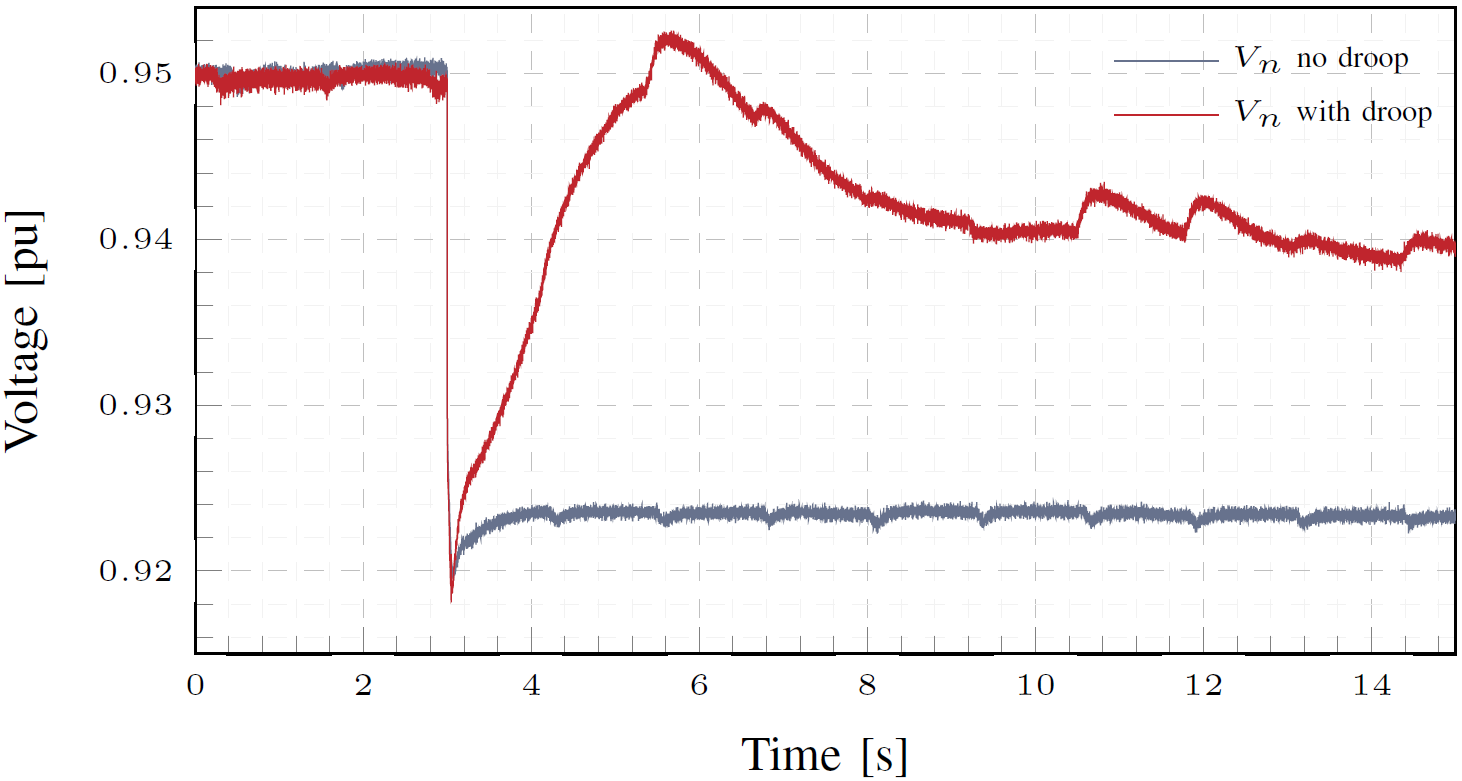}
\caption{RMS voltage response at the point of common coupling on the grid side following a load increase by \SI{1}{\mega\watt} with and without droop control of the BESS system}\label{fig:rms_voltage}
\end{figure}

\begin{figure}[!t] 
\centering
\includegraphics[width=0.75\textwidth]{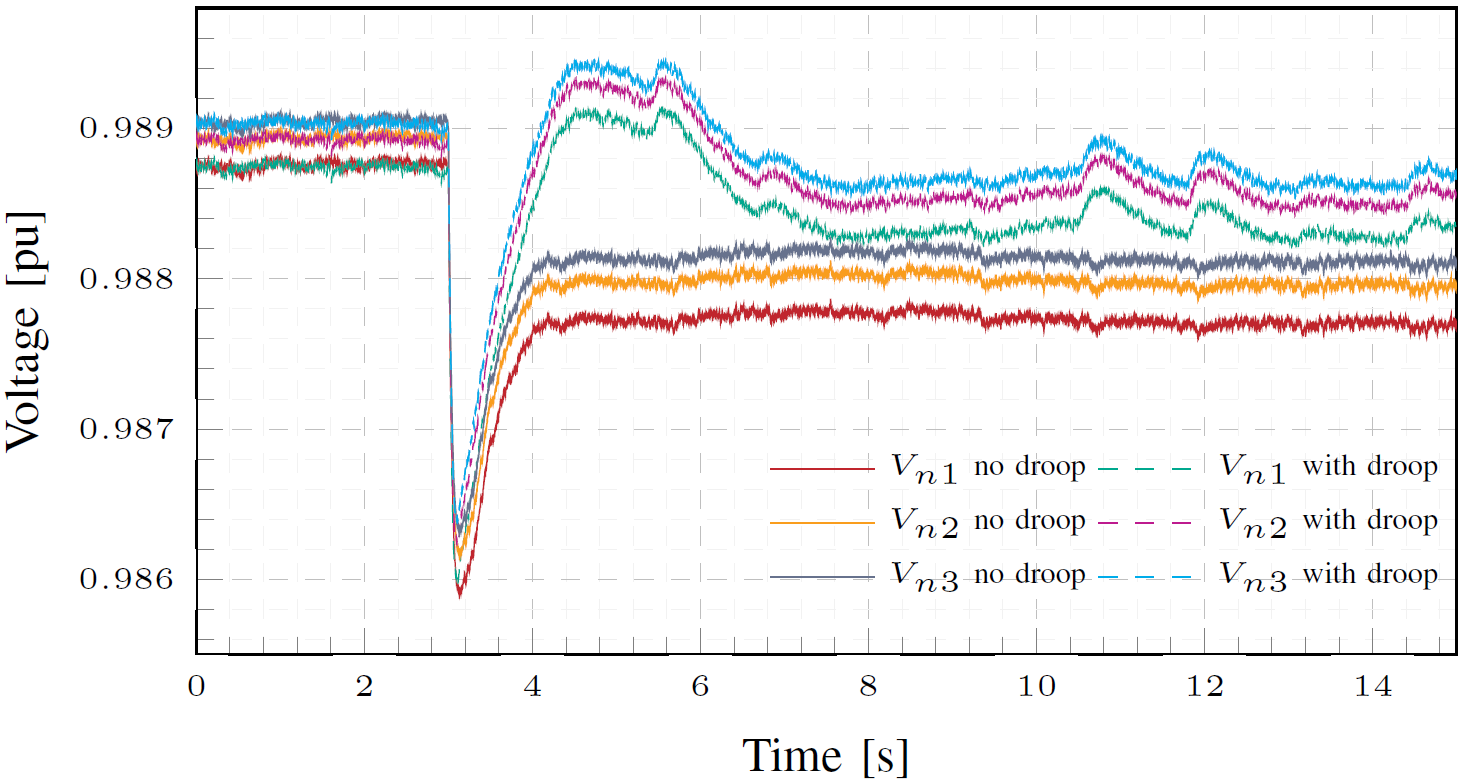}
\caption{RMS voltage response at additional measurement points close to the point of common coupling on the grid side}\label{fig:rms_voltage_distnodes}
\end{figure}

The RMS voltage at the point of common coupling on the grid side is shown in Fig.~\ref{fig:rms_voltage}. It can be observed in Fig.~\ref{fig:rms_voltage} that activation of the droop control strongly supports the network voltage at the point of common coupling. 
Furthermore, additional measurement points are considered in order to investigate the system wide influence of the BESS system on the grid voltage. The selected measurement points are three consecutive stations moving away from the point of common coupling (i.e. $n_{1}$, $n_{2}$ and $n_{3}$, cf. Fig.~\ref{fig:fig_KITCN}). Fig.~\ref{fig:rms_voltage_distnodes} shows the corresponding RMS voltages at the additional measurement points. Comparing the cases with and without droop control in Fig.~\ref{fig:rms_voltage_distnodes}, it can be observed that the voltage support provided by the BESS droop control is also experienced at other nodes in the network. This shows that the experimental setup can be used to study the potential of microgrid control on system wide grid support in response to network disturbances. 


\subsection*{Unbalanced Load Change in Microgrid}\label{subsec:5_exp2_3}
\begin{figure}[!t] 
\centering
\includegraphics[width=0.75\textwidth]{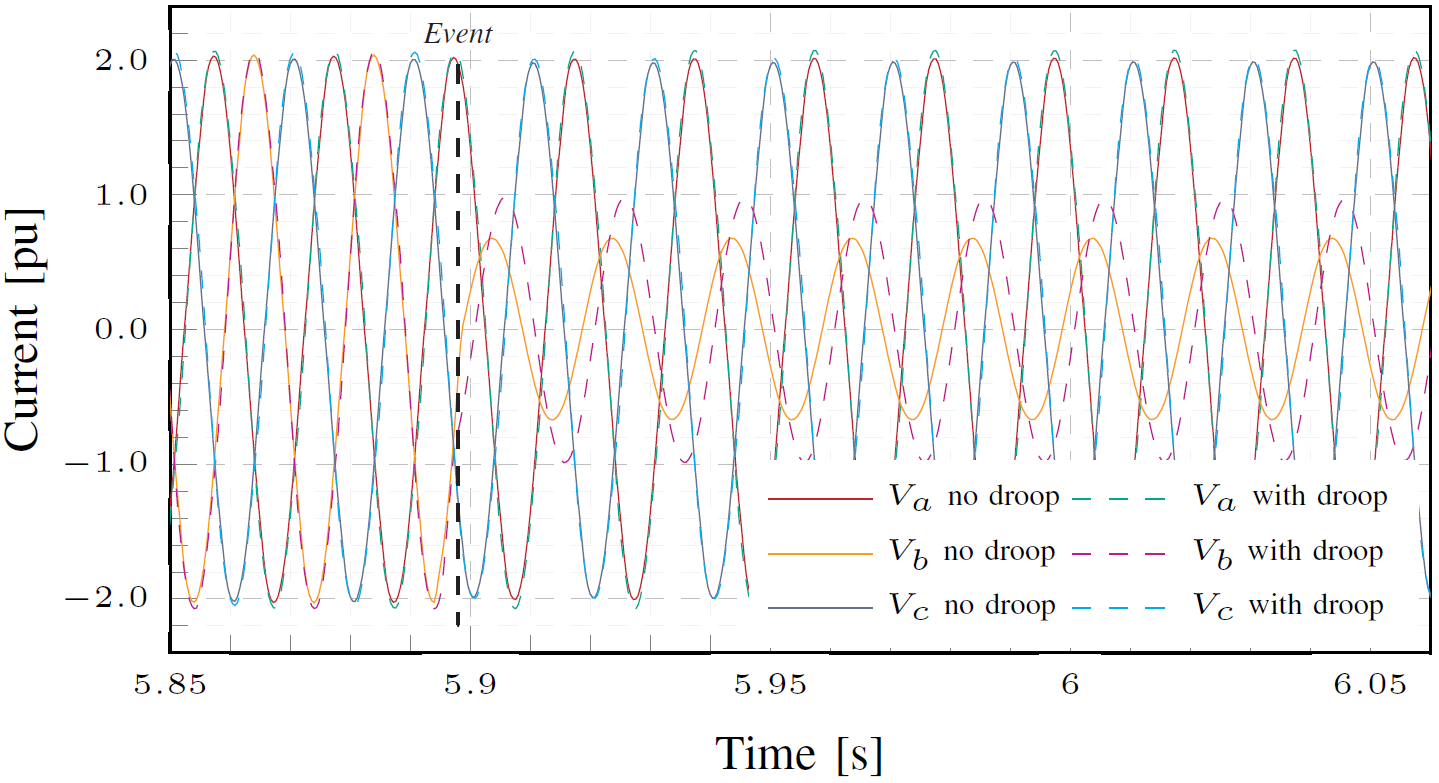}
\caption{Measured instantaneous current at point of common coupling on the network side following an unbalanced  decrease in load on phase B }\label{fig:inst_currents_exp2}
\end{figure}

The second case tests the response on the simulated network side following an event on the hardware side of the microgrid experiment. For this, an unbalanced load is triggered on the resistive load hardware in the microgrid setup in order to analyze if the effect is detected in the simulated network. The first load change is a decrease in the resisitive load from \SI{20}{\kilo\watt} to \SI{0}{\kilo\watt} on phase B. 
Fig.~\ref{fig:inst_currents_exp2} shows the response of the measured instantaneous currents on all three phases at the point of common coupling on the network side for the cases with and without droop control setting in the BESS system.
Important to observe is that the unbalanced load effect is experienced on the network side, since the measured current on phase B defers in amplitude compared to the other phases after the event is triggered. 
In addition, it can be observed that the droop control setting of the BESS system influences the current change with an unbalanced event in the experimental setup. 


 
\begin{figure}[!t] 
\centering
\includegraphics[width=0.75\textwidth]{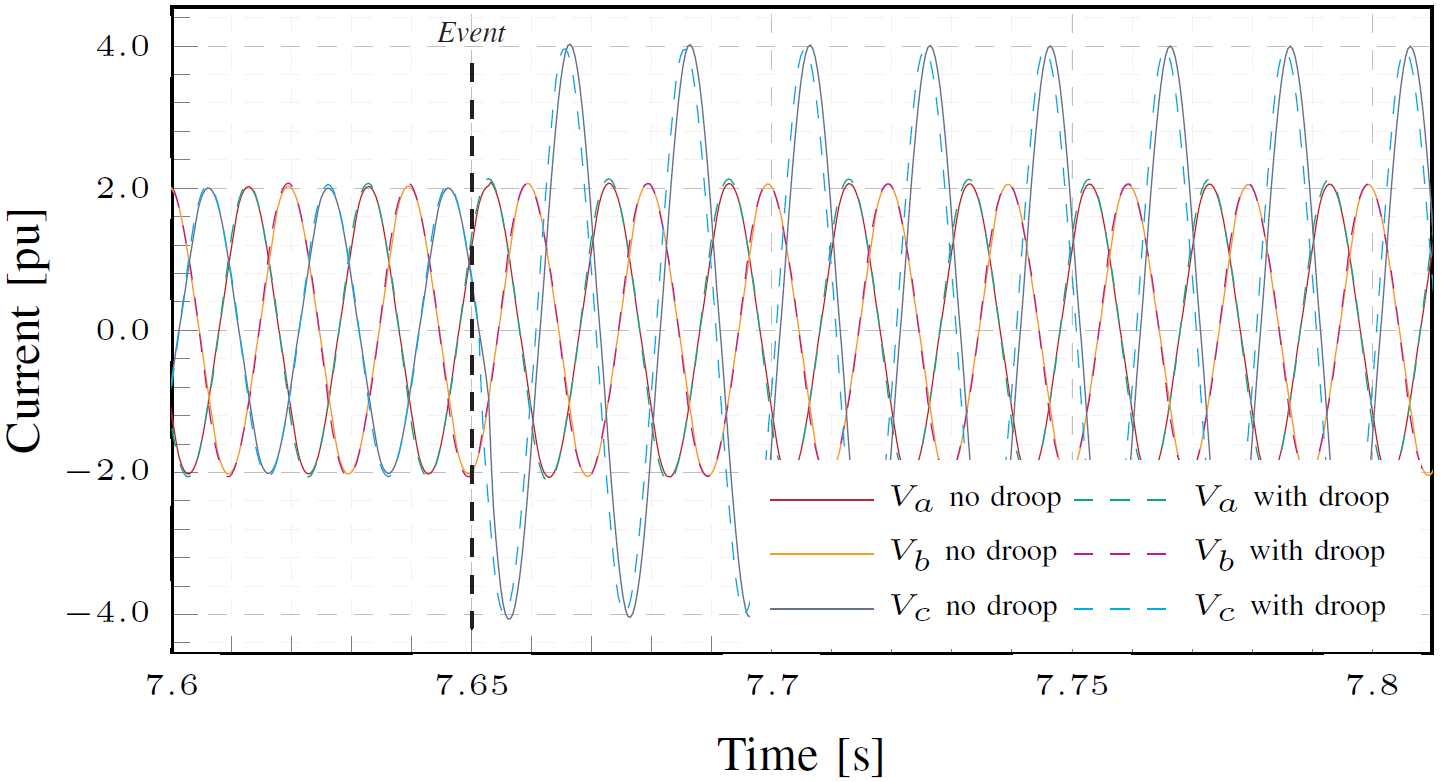}
\caption{Measured instantaneous current at point of common coupling on the network side following an unbalanced  increase in load on phase C }\label{fig:inst_currents_exp3}
\end{figure}

The second unbalanced load change scenario is an increase in the resistive load from \SI{20}{\kilo\watt} to \SI{60}{\kilo\watt} on phase C. Fig.~\ref{fig:inst_currents_exp3} shows the unsymmetrical measured currents at the point of common coupling. The current in phase C increases in response to the increase in load as observed in Fig.~\ref{fig:inst_currents_exp3}. 
This therefore shows that the simulated network can correctly capture the unsymmetrical behavior in the interfaced experimental hardware. 


 
\section{Conclusion and Outlook}\label{sec:6conclusion} 

The present article introduces the physical interface between hardware-based microgrid experiments and real-time simulated power grids using the Power Hardware-in-the-Loop approach. This setup demonstrates the advantages of combining physical experiment hardware and digital representations of large grids in testing microgrid control strategies for future energy systems. An important observation in the presented setup is that the communication delays via the physical interface do not significantly affect the accuracy in the interaction between the physical hardware and the simulated network. The communication delay is observed to depend only on the simulation sampling time. 
It is also shown that possible signal distortions can have a visible influence on the quality of the results.
Moreover, the results of the experimental studies show that the setup can be reliably used to test microgrid control strategies and their effect in supporting the main grid in response to system wide disturbances.   

Future work will focus of including additional physical power system hardware components in the microgrid experimental setup. In this way, further strategies, like control and coordination of distributed energy resources will be implemented in order to test coordinated grid support functionality in a fully hardware-based microgrid. Furthermore, the setup will be extended to include hardware from distant energy research infrastructures through a virtual connection framework for geographically distributed real-time simulations.

\section*{Acknowledgment}
This work was supported by the Helmholtz Association under the program ``Energy System Design".

\bibliographystyle{unsrtnat}

\end{document}